\documentstyle[prl,aps]{revtex}
\input{epsf.sty}

\def\b\mu{{\bf \mu}}

\begin{document}

\draft

\title{ Spreading of Wave Packets, Uncertainty Relations and the \\
de Broglie Frequency }
\author{H. C. G. Caldas$^{\dagger}$$^{\ddag}$\thanks{e-mail: hcaldas@funrei.br}, 
and P. R. Silva$^{\ddag}$}

\address{${\dagger}$Departamento de Ci\^{e}ncias Naturais, DCNAT \\
Funda\c{c}\~ao de Ensino Superior de S\~{a}o Jo\~{a}o del Rei, FUNREI,\\ 
Pra\c{c}a Dom Helv\'ecio, 74, CEP:36300-000, S\~{a}o Jo\~{a}o del Rei, MG, Brazil} 

\address{${\ddag}$Departamento de F\'\i sica, Universidade Federal de Minas Gerais,\\
CP 702,CEP:30.161-970,Belo Horizonte, MG, Brazil}

\maketitle                          

\begin{abstract}
The spreading of quantum mechanical wave packets is studied in two
cases. Firstly we look at the time behavior of the packet width of a
free particle confined in the observable Universe. Secondly, by imposing the
conservation of the time average of the packet width of a particle driven by a harmonic oscillator potential, we find a zero-point energy which frequency
is the de Broglie frequency.

\end{abstract}


\newpage 
 

The quantum mechanical wave-packet spreading is a subject of current
interest as can be verified in some recently published papers \cite{grobe}
and \cite{dodo}. As pointed out by Grobe and Fedorov \cite{grobe} the
ionization of atoms can be supressed in superstrong fields. This phenomenon
has been called stabilization and is characterized by decreasing ionization
probability with increasing laser intensity. The wave-packet spreading plays
a key role in the final degree of stabilization. On the other hand, Dodonov
and Mizrahi \cite{dodo} have adressed to the ``Strict lower bound for the
spatial spreading of a relativistic particle`` where they provide a strict inequality for the minimal possible extension of a wave packet corresponding to the physical state of a relativistic particle.

In this letter we intend to study the spreading of wave-packets in two
particular situations. In the first case we explore the consequences of the
finiteness of the Universe in the spreading of a free particle wave-packet.
In the second one, we want to study the wave-packet of a particle described
by an one-dimensional harmonic oscillator. As we will see this can lead to
interesting consequences related to the interpretation of the de Broglie
frequency of a particle.
For a one-dimensional wave-packet let us define \cite{messiah}

\begin{eqnarray}
\left( \Delta q\right) ^2 &=&\left\langle q^2\right\rangle -\left\langle
q\right\rangle ^2 \,,  \label{um} \\
\left( \Delta p\right) ^2 &=&\left\langle p^2\right\rangle -\left\langle
p\right\rangle ^2 \,,  \nonumber
\end{eqnarray}
Where $\left( \Delta q\right) ^2$ and $\left( \Delta p\right) ^2$ in the above relations are, respectively the variancies of the quantities q and p, representing the position and the momentum of a particle. An interesting interpretation of wave-packet spreading can be
found in Gasiorowicz \cite{gaz}.

As pointed out by Messiah \cite{messiah}, the spreading law for a free
wave-packet turns out to be quite simple if the wave packet is taken to be the minimum at the initial time, namely:

\begin{equation}
\Delta q_0\Delta p_0=\frac 12\hbar \,. \label{dois}
\end{equation}

Besides this, the Heisemberg equation of motion gives in this case:

\begin{equation}
\left( \Delta q\left( t\right) \right) ^2=\left( \Delta q_0\right) ^2+\frac{%
\left( \Delta p_0\right) }{m^2}^2t^2 \,.  \label{tres}
\end{equation}

Now, looking at relation (\ref{tres}) we verify that a lower bound for the
spreading in the particle localization corresponds to a lower bound on the
initial (t=0) variance of the particle momentum
$\left( \Delta p_{0} \right) $. Considering the universe finiteness this result is just an approximation (certainly a very good one) of taking as free a wave packet confined to a very large box. Can we associate this $\left( \Delta p_{0}
\right) $ minimum with the finiteness of the Universe? In the folowing we
are going to look for this possibility.

In his paper: "Is the Universe a Vacuum Fluctuation?" Tryon \cite{Tryon} states that the positive mass energy of a particle could be cancelled by an equal amount of negative gravitational energy, due to the interaction of this particle with the rest of the Universe. In this way the classical mechanical energy of a particle is equal to zero, so that in average the particles are free. However, from the point of view of the quantum mechanics, we may permit fluctuations in the energy of this particle. Besides finite the Universe has also a finite age and this implies that we cannot have a complete freedom for preparing an wave packet with arbitrary initial conditions. So let us consider the uncertainty in time to be of the order of the Hubble time $H_{0} ^{-1}$. Now, we write the minimum time-energy uncertainty relation, namely: 

\begin{equation}
\Delta E\Delta t=\frac 12\hbar \,. \label{quatro}
\end{equation}
Putting $\Delta t=H_{0}^{-1}$ in (\ref{quatro}), we obtain:

\begin{equation}
\Delta E=\frac 12\hbar H_0 \equiv E_1\,.  \label{cinco}
\end{equation}

We observe that the Hubble constant is related to the radius of the Universe through the equation

\begin{equation}
H_{0} =\frac cR_{0} \,. \label{seis}\end{equation}
Therefore the lower bound on the kinetic energy of a particle reflects the fact that the Universe has a finite radius.
We assume, on cosmological grounds, that each particle has in average zero total (mechanical) energy, but with fluctuations about this zero energy.

Now we supose that the minimum uncertainty in the kinetic energy of a
particle confined in the Universe corresponds to this lowest energy level $E_1$
and we write the equality:

\begin{equation}
\frac{\left( \Delta p_0\right) }{2m}^2=E_1=\frac 12\hbar H_0 \,.  \label{doze}
\end{equation}

By using the relation $\Delta p_0=m\Delta v_0$ , and after solving for $%
\Delta v_0$ , we get:

\begin{equation}
\left( \Delta v_0\right) _{min}=\sqrt{\frac{\hbar H_0}m} \,.  \label{treze}
\end{equation}

The minimum uncertainty in velocity given by (\ref{treze}) can be
interpreted as a lower bound on a particle velocity. For the electron we have
$\left( \Delta v_{0} \right) _{min} =4 \cdot 10^{-12} \frac{m}{s} $.

From (\ref{doze}), we also have:

\begin{equation}
\left( \Delta p_0\right) =\sqrt{m\hbar H_0} \,.  \label{quatorze}
\end{equation}

Putting (\ref{quatorze}) into (\ref{dois}) ( the minimum uncertainty
relation ) we obtain

\begin{equation}
\left( \Delta q_0\right) =\frac 12\sqrt{\frac \hbar {mH_0}}=\frac 12\sqrt{{\lambda}_{rc}R_0} \,,  \label{quinze1}
\end{equation}
where $R_{0}=\frac cH_0 $ is the radius of the Universe and ${\lambda}_{rc}=\frac \hbar {mc}$ is the reduced Compton wavelength
of the particle. Then we see that, for the minimum uncertainty in the
momentum, the particle has an uncertainty in position which is the geometric
average between ${\lambda}_{rc}$ ( a characteristic length of the
particle ) and $R_0$ ( the radius of the observable Universe ). Eq. (\ref{quinze1}) is the output from a compromise of two contradictory conditions of free evolution and of being hold inside a closed box. Despite this
maximum initial uncertainty in the position of the particle being very large ( it is of the order of $10^6m$, for the electron ) it is, for the electron, $%
10^{20}$ times smaller than the radius of the Universe. This maximum
initial variance of a physical coordinate of a particle coupled to the Universe must be compared with its minimum \cite{dodo} which is given by $\frac 12{\lambda}_{rc}$.

Eq.(\ref{quinze1}) also deserves the following comment. Thinking in terms of a hyphotetical Universe where $R$ is a variable quantity, we can write:
\begin{equation}
\left( \Delta q_0\right) =\frac 12\sqrt{{\lambda}_{rc}R} \,.\label{quinze}
\end{equation}
In this way the initial variance of a particle position will be able to vary from its minimum to its maximum value, with R running from ${\lambda}_{rc}$ to $R_0$. 
   
Now, if we consider that the initial time corresponds to the present time,
let us see what happens with the spreading of an wave packet with initial uncertainty given by Eq. (\ref{quinze1}) and if we wait a time equal to the Hubble time ( $t=H_{0} ^{-1} $ ). Using (\ref{tres}) we get:

\begin{equation}
\left( \Delta q\left( t_H\right) \right) =\sqrt{\frac 54}\sqrt{\frac \hbar
{mH_0}}=\sqrt{5}\left( \Delta q_0\right)  \label{dezeseis}
\end{equation}
In obtaining (\ref{dezeseis}), we also used (\ref{quatorze}) and (\ref
{quinze1}).

Therefore we verify that if we wait a time equal to the Hubble time, the
initial maximum variance of a physical coordinate of a particle is not substantially modified.

We will now look to the second case when we study the spreading of a wave-packet of a particle described by a one-dimensional harmonic oscilator.

As pointed out by Messiah \cite{messiah}: In order that the motion of a wave
packet may be likened to the motion of a classical particle, it is first of
all necessary that its position and momentum follow the laws of classical
mechanics. Also according to Messiah the two most interesting cases are
those of the harmonic oscillator and the free particle, cases for which the
motion of the center of the packet is rigorously identical to that of a
classical particle. Let us turn now our atention to the harmonic oscillator
case. In a pedagogical paper \cite{perez} the Heisenberg representation was
used as a mean to study the spreading of wave-packets in some simple
examples. For the harmonic oscillator, whose hamiltonian is given by:

\begin{equation}
H=\frac{p^2}{2m}+m\omega ^2\frac{q^2}2  \,, \label{dezesete}
\end{equation}
the evolution in time of the width in the position distribution is given by 
\cite{perez}:
\begin{equation}
\left( \Delta q\left( t\right) \right)^2=\left( \Delta q_0\right) ^2\cos
^2\left( \omega t\right) +\frac{\left( \Delta p_0\right) }{\left( m\omega\right) ^2}^2sin^2\left( \omega t\right) +\left( \frac 12\left\langle
qp+pq\right\rangle _0-\left\langle q\right\rangle _0\left\langle
p\right\rangle _0\right) \frac{sin\left( 2\omega t\right) }{m\omega } \,.
\label{dezoito}
\end{equation}

Now let us make the requirement that 

\begin{equation}
\left[ \Delta q\left( t\right) ^2\right]_{time\,average}=\frac{%
\left( \Delta q_0\right) }2^2+\frac 12\frac{\left( \Delta p_0\right) }{%
\left( m\omega \right) ^2}^2=\left( \Delta q_0\right) ^2  \,, \label{dezenove}
\end{equation}

where we have averaged $\left( \Delta q\left( t\right) \right) $ in a period
of time equal to $T=\frac{2\pi }{\omega } $ .
Relation (\ref{dezenove}) implies that:

\begin{equation}
\left( \Delta p_0\right) =m\omega \left( \Delta q_0\right) \,.  \label{vinte}
\end{equation}
Putting (\ref{vinte}) into the minimum uncertainty relation (\ref{dois}), we
obtain:

\begin{equation}
m\omega \left( \Delta q_0\right) ^2=\frac 12\hbar \,. \label{vinteum}
\end{equation}
Multiplying both sides of (\ref{vinteum}) by $\omega $ , we get:

\begin{equation}
m\omega ^2\left( \Delta q_0\right) ^2=\frac 12\hbar \omega \,. \label{vintedois}
\end{equation}

On the other hand, for a classical harmonic oscillator of amplitude A, we can write:

\begin{equation}
q\left ( t\right)_{class} =A cos\left(\omega t\right) \,.
\label {vintetres}
\end{equation}
The above relation leads to:

\begin{equation}
\left[ \Delta q\left( t\right)_{class} ^2\right]_{time\,average}=\frac {A^2}2\,.
\label{vintequatro}
\end{equation}
Making the requirement that the classical variance to be identified with the quantum variance $\left ( \Delta q_0\right)$, we obtain  

\begin{equation}
m\omega ^2 \left( \Delta q_0\right) ^2=\frac 12m\omega ^2 A ^2=\frac 12\hbar\omega\,.
\label {vintecinco}
\end{equation}
Second and third terms of equation (\ref{vintecinco}) show a classical harmonic oscillator which mechanical energy is equal to the zero-point energy of the corresponding quantum oscillator.

An interesting consequence of relation (\ref{vintecinco}), despite its non-relativistic character, is obtained when we take the speed of light $c$ as a limit for the velocity of the particle undergoing classical harmonic motion. Putting $\omega A=c$ in equation (\ref{vintecinco}), we get
 
\begin{equation}
mc^2=\hbar \omega \equiv \hbar \omega _{dB} \,. 
\label{vinteseis}
\end{equation}
and

\begin{equation}
A= \frac \hbar{mc} \,. \label{vintesete}
\end{equation}

Therefore we see that (\ref{vinteseis}) reproduces the definition of the de Broglie frequency implying also that the classical amplitude of the oscillator to be equal to the reduced Compton wavelength. The driving force amplitude of this oscillator is given by:

\begin{equation}
F_1 = m\omega^2A=\frac {m^2c^3}\hbar \,.
\label{vinteoito}
\end{equation}

It can also be interpreted as a string constant. Some numerical estimates of it gives order of magnitudes of $10^{-1}N$ for the electron and $10^5N$ for the nucleons (protons or neutrons). The fact that the force $F_1$ is proporcional to the squared mass of the particle and that it can be defined for electrons, protons, neutrons or any other kind of elementary particles, lead us to think in the only common kind of interaction experimented by these various particles, namely: the gravitational interaction. If we multiply and divide equation 
(\ref{vinteoito}) by $G$ (the gravitional constant), we obtain:

\begin{equation}
F_1=\frac {Gm^2}{\lambda_P\,^2} \,.
\label{vintenove}
\end{equation}
where 

\begin{equation}
\lambda_p\,^2 =\frac {G\hbar}{c^3} \,.
\label{trinta}
\end{equation}
is the square of the Planck radius.

In conclusion, we would like to take into account the following
considerations \cite{anandam}: It was pointed out by Penrose \cite{penrose}
that the existence of accurate clocks is ultimately due to the fact that
each particle of mass m has associated with it, a natural frequency
$\omega _{dB} $ given by the Einstein-Planck's law $\Delta E=mc^{2} =\hbar
\omega _{dB} $ .

Therefore we can associate this natural frequency to the de Broglie
frequency, with the driving force behind this clock being attributed to the
internal degree of freedom of the particle described by a harmonic
oscillator potential. The same conclusion was reached by one of the present authors \cite{paulo} starting from other initial assumptions.

{\small 
}

\subsection*{Acknowledgements}

{\small 
CNPq-Brazil is acknowledged for invaluable financial help. We are grateful to Dr. A. L. Mota and Dr. Manoelito for critical reading of the manuscript and for discussions. }

{\small 
}

\newpage



\begin{references} 

\let\ul=\underbar
\def\AP#1{{Ann. Phys. }{\bf #1}}
\def\PRP#1{{Phys. Rep. }{\bf #1}}
\def\APP#1{{Act. Phys. Pol. }{\bf #1}}
\def\PTP#1{{Prog. Theor. Phys. }{\bf #1}}
\def\PR#1{{Phys. Rev. }{\bf #1}}
\def\PRD#1{{Phys. Rev. }{\bf D #1}}
\def\PRC#1{{Phys. Rev. }{\bf C #1}}
\def\PRL#1{{Phys. Rev. Lett. }{\bf #1}}
\def\PL#1{{Phys. Lett. }{\bf #1}}
\def\RMP#1{{Rev. Mod. Phys.}{\bf #1}}
\def\NP#1{{Nucl. Phys. }{\bf #1}}
\def\ZP#1{{Z. Phys. }{\bf #1}}
\def\NC#1{{Nuovo Cimento }{\bf #1}}
\def\SJNP#1{{Sov. J. Nucl. Phys. }{\bf #1}} 
\bibitem{grobe}  {\small {R. Grobe and M.F. Fedorov, {\em {Phys.Rev.Lett.}} 
{\bf {68}} (1992) 2592.} }

\bibitem{dodo}  {\small {V.V. Dodonov and S.S. Mizrahi, {\em {Phys.Lett.A}} 
{\bf {117}} (1993) 394.} }

\bibitem{messiah}  {\small {A.Messiah, {\em {Quantum Mechanics, Vol I, John
Wiley and Sons,}} N. York, 1958.} }

\bibitem{gaz}  {\small {S.Gasiorowicz, {\em {Quantum Physics, John Wiley,}}
N. York, 1974.} }

\bibitem{Tryon}  {\small {E. P. Tryon, {\em {Nature}} {\bf {246}} (1973) 396.} }

\bibitem{perez}  {\small {J.F. Perez, {\em {Revista Brasileira de Ensino de
Fisica}} {\bf {17}} (1995) 123.} }

\bibitem{broglie}  {\small {L. de Broglie, {\em {Ann. Phys.}} {\bf {2}}
(1925) 10}, see also {J.W. Haslett, {\em {Am.J. Phys.}} {\bf {40}} (1972)
1315.} }

\bibitem{anandam}  {\small {J.S. Anandam, {\em {Quantum Theory of
Gravitation Ed. A.R. Marlow, Proc. of Simposium held at Loyola University,
New Orleans, Academic Press,}} 1980.} }

\bibitem{penrose}  {\small {R. Penrose , {\em {In Battelle Recontres,Eds.
C.M. De Witt and J.A. Wheeler, Benjamin, N.York,}} 1968.} }

\bibitem{paulo} {\small {P.R. Silva, {\em {Phys. Essays}} {\bf {10}} (1997) } }

\end{references}
\end{document}